\documentclass{article}
\usepackage{amsmath}
\usepackage{amscd}
\usepackage{amsthm}
\usepackage{amssymb} \usepackage{latexsym}
\usepackage{eufrak}
\usepackage{euscript}
\usepackage{epsfig}
\usepackage{graphics}
\usepackage{array} 
\usepackage{enumerate}
 
\usepackage{pictexwd,dcpic}

\usepackage{color}

\usepackage{hyperref}

\newcommand{\bel}[1]{\begin{equation}\label{#1}}

\newcommand{\be}{\begin{equation}}

\newcommand{\ba}{\begin{eqnarray}}
\newcommand{\ea}{\end{eqnarray}}
\newcommand{\rf}[1]{(\ref{#1})}
\newcommand{\bi}{\bibitem}
\newcommand{\qe}{\end{equation}}

\newtheorem{thesis}{Thesis}
\newcommand{\btl}[1]{\begin{thesis}\label{#1}}
\newcommand{\et}{\end{thesis}}

\theoremstyle{theorem}

\theoremstyle{corollary}

\theoremstyle{lemma}

\theoremstyle{definition}

\theoremstyle{proof}

\theoremstyle{remark}

\usepackage{tikz-cd} 

\usepackage{tikz}

\title{Biological information}
\author{J\"urgen Jost}
\begin{document}
\date{}
\maketitle

\begin{abstract}
  In computer science, we can theoretically neatly separate transmission and processing of information, hardware and software, and programs and their inputs. This is much more intricate in biology, Nevertheless, I argue that Shannon's concept of information is useful in biology, although its application is not as straightforward as many people think. In fact, the recently developed theory of information decomposition can shed much light on the complementarity between coding and regulatory,  or  internal and environmental information. The key challenge that we formulate in this contribution is to understand how genetic information and external factors combine to create an organism, and conversely, how the genome has learned in the course of evolution how to harness the environment, and analogously, how coding, regulation and spatial organization interact in cellular processes. 
\end{abstract}

\section{Introduction}
Fundamental scientific concepts seem to have the tendency to give rise to much confusion, and the concept of information certainly is no exception. 

Remarkably, however, the  concept of information is relatively uncontroversial in computer science. Its application in quantum physics created  quantum information theory. The confusions of the two most prominent opponents on the foundations of quantum physics, Bohr and Einstein, have been overcome, and in particular, the objections of Bohr have been conclusively refuted by experiments confirming the phenomenon predicted by Einstein-Podolsky-Rosen \cite{EPR} if quantum theory were right,  but considered as so absurd by them as to refute quantum theory. To clarify the issue, Schr\"odinger introduced the concept of entanglement. As a consequence, quantum information theory is now a striving field with the potential for important technological applications.  

In other fields like biology or cognition, however,  the  debates and controversies about the  concept of information have not yet been concluded. Some of these concern the questionable applicability of metaphors from computer science in those domains, and others simply come from misunderstandings. And philosophers who tried to enter the debates did not always make the situation clearer. 

Of course, several people did come up with positive insight and suggested clarifications, but unfortunately, not all of that has been generally accepted so far. 

It may therefore not be completely useless to develop a systematic analysis of biological information. That is what I shall attempt here. The role of information for understanding cognition will be discussed elsewhere, so that here I can concentrate on biology. 

A small technical point: I shall mainly speak about organisms, that is multicellular eukaryots. I shall also repeatedly invoke the example of viruses. Obviously, what I shall say will also apply mutatis mutandis to prokaryots.

\section{Information}\label{info}
In computer science, the widely accepted formal definition of information is that introduced by Claude Shannon \cite{Sha}. In abstract terms, it treats the situation where we have  a random variable $X$, or equivalently,  a probability distribution $p$, when the possible values $x_i$ of $X$ are realized with probabilities $p_i=p(x_i)$. As probabilities, they have to  satisfy $0\le p_i \le 1$ for all $i$ and  the normalization $\sum_i p_i=1$. The {\em Shannon information}  then is
\bel{1}
H(X)=H(p_1,\dots, p_n)=-\sum_i p_i \log  p_i \text{ (bits)}.
\qe
We shall always use binary logarithms, that is, $\log=\log_2$, which is
Shannon's convention. 
$H(X)$ is also called the {\em Shannon entropy}, because of deep analogies with the concept of entropy in statistical mechanics, but here, we do not go into that. Nevertheless, we shall speak of an entropy, and this may suggest a certain interpretation to some readers.\\
While the formula \rf{1} is as clear as it could possibly be, its genesis was not easy  and its interpretation has caused much confusion. For instance, the founder of cybernetics, Norbert Wiener, had tried to push a different notion. 

The correct interpretation, however, simply is the following. 
$H(X)$   is the expected {\em reduction of uncertainty}, i.e., the {\em information gain}, if we learn which concrete value $x_i$ of the random variable  $X$ from a known distribution  $p$ with probabilities  $p_i=p(x_i)$ is realized. \\
Here is the basic example: When we have two possible events occurring with equal probability 
 1/2 (an unbiased coin) we thus gain  $\log 2=$1 bit (recalling the convention
 $\log =\log_2$) of information when we observe the outcome. When we observe the result of a  fair dice, we obtain $\log  6$ bits of information.  \\
In fact, the normalization just employed, that by observing one of two equally likely events, we obtain 1 bit of information, together with three other axioms, continuity, monotonicity and additivity, uniquely determine $H(X)$ in \rf{1}, and this was the axiomatic approach of Shannon. Continuity here means that $H(X)$ should depend continuously on the $p_i$, monotonicity that $H(X)$ is the larger the more equally likely events we have, and additivity that when we gain the information in subsequent steps, the total gain should equal the sum of that obtained from those steps. Thus, when we learn the result of the dice in two steps, first whether the result is $\le 3$ or $>3$ ($\log 2=$1 bit of information since the two possibilities are equally likely) and then the precise value (remaining uncertainty $\log  3$ bits, since in either case there are three possible numbers), we altogether get $\log  2+ \log  3=\log  6$ bits.  (While probabilities multiply, the logarithm in \rf{1} makes information additive.) \\
Again, at least when one understands some basic mathematics, this seems as
clear as it could possibly be. But when trying to interpret it, problems
emerge. Actually, Claude Shannon was working at Bell Labs, that is,  for a
telephone company, and he was interested in how much information  per unit of
time could be transmitted along a phone line. Let us simplify the situation
and consider letters converted into Morse code and transmitted along a
telegraph cable. Thus, we have a sender who encodes his symbols, the letters,
in the code, a channel through which the coded signals  are transmitted, and a
receiver who can then decode the signals back into letters. The transmission
along the cable might be perturbed by noise. The capacity is limited, because
the sender can send the Morse symbols only at some finite speed.  The original
symbols (26 letters plus a stop sign) have different relative frequencies in
English text. The relative frequency with which letter $i$ occurs then becomes
its   probability $p_i$ in \rf{1}.  $H(X)$ then is 
the average number of bits transmitted per symbol. In order to optimize the
transmission, we should  use  short representations for
frequent letters. That is, letters with a high probability, like e or r,
should be represented by a short code word. Since the number of short code
words is limited, for less frequent letters,
like x, we could take  longer code words. Thus, the average length of a code
word should be as small as possible to make the code efficient.  If on the other hand, one wants to reduce the ambiguities in the decoding process caused by channel noise, the code has to  become redundant and thereby less efficient. Shannon's theory easily and beautifully resolves these issues. \\
Obviously, this can be formulated in abstract terms. We have a sender who sends signals over a channel to a receiver, and when the receiver receives them he knows what they stand for and therefore can understand the message. Before receiving this message, he only knew the probabilities of the symbols, but not which precise ones the sender would choose. \\
Let us take a closer look at the underlying assumptions. Sender and receiver both need to know the code that is utilized, the Morse code in our examples. Thus, when the receiver receives a string of signals, he knows which letters they stand for, and in the absence of noise, he can therefore unambiguously decode the message. In particular, sender and receiver have to share some prior knowledge. \\
Also, the preceding applies to any string of letters translated into Morse code and transmitted via the channel. It does not take the fact into account that an English text is transmitted. In English, not all combinations of letters are permitted, and in turn, certain combinations, like {\em wh},  are more likely than the probabilities of the individual letters suggest.\\
 Before proceeding, we need to clarify an important point. A message typically
 does not consist of a single symbol, but it is a string of symbols whose
 composition is constrained by certain rules. When  the sender transmits a
 random sequence of symbols, with symbols chosen independently of previous
 ones, and if the entropy per symbol is $H$ and the length of the sequence is
 $n$, then by the additivity of Shannon entropy, the entire sequence has
 entropy $nH$. When, however, the symbols in the sequence are not entirely
 independent of each other, for instance when the letters have to form English
 words, the entropy becomes smaller. Of course, we could then take English
 words instead of letters from the Latin alphabet as our basic symbols and
 evaluate the entropy of the message in terms of their probabilities. Again,
 there might be higher order correlations, because in a text, the words are
 not chosen randomly, but their composition has to follow grammatical and
 stylistic rules. This further reduces the uncertainty contained in the text,
 because having received some words, we can often already guess the next one,
 so that there is no further uncertainty. And if we know what type of text to
 expect, we are even less uncertain about the actual message. That is, the
 receiver may possess additional prior knowledge about the text that makes his
 uncertainty before receiving it smaller. That is, in Shannon's sense, he will
 then obtain less information from receiving the text. As explained, the
 information contained in an arbitrary and probably completely meaningless
 string of symbols, all chosen with equal probability, is higher than that
 obtained from any structured English text. That may sound
 counterintuitive. In particular, there does not seem to be  any  room for the
 meaning of the transmitted, even though colloquial use of the word {\em
   information} would suggest that a meaningful text contains much more of it than meaningless ones. \\
As a possible solution, one might want to quantify the value of the information received, in economics by how much you would be willing to pay for it, or in biology by your fitness gain from being able to exploit it, for example from learning where food or a mate can be found. But that would constitute an additional principle. From a more intrinsic perspective, the receiver can choose, or the sender and the receiver can agree, on what there is uncertainty to be reduced by transmitting and attending to a signal. We receive lots of sensory input all the time, but we only attend to some of it. When the input is just random, and if we already know that it is random, there is no uncertainty to be reduced. We may bring in Bateson's slogan \cite{Bat} of information as a {\em difference that makes a difference}. That is, what counts as information depends on a receiver and what he entertains uncertainty about. Whatever the sender sends, if nobody listens to her, there is no information. And if it simply looks random to a receiver, there is no information either. Only when the receiver gets interested in the particular symbols, and if these are drawn from a known probability distribution, there is information and what the  sender emits becomes a  message. While this does not yet fully resolve the issue of meaning, hopefully it can at least refute some misconceptions. \\
Before proceeding, however, we need to clarify another important issue. It seems that \rf{1} defines some absolute quantity. But we had already emphasized that information is a reduction of uncertainty. That means that we evaluate the difference between what the receiver knows before and after receiving the message. In other words, the information depends on the prior state of knowledge of the receiver. It is therefore not an absolute quantity, but a relative one. In order to formalize that, we need the concept of conditional information. \\
Thus, in addition to $X$, we have another random variable $Y$ that can assume values $y_j$ with probabilities $q_j$. The conditional probability $p(x_i|y_j)$ is the probability that $X$ assumes
the
value $x_i$ given that $Y$ has the value $y_j$. We have 
\begin{equation}
\label{2}
p(x_i)= \sum_j p(x_i|y_j) q(y_j).
\end{equation}
We also have the joint probability 
\begin{equation}
\label{3}
p(x_i,y_j)=p(x_i|y_j) q(y_j)
\end{equation}
that the values $x_i$ and $y_j$ occur together. From \rf{2} and \rf{3}, we see that things are consistent, that is, we have
\bel{4}
p(x_i)= \sum_j p(x_i,y_j),
\qe
expressing $p(x_i)$ as a so-called marginal probability. \\
The conditional entropy of $X$ given $Y$ then is defined as
\begin{equation}
\label{5}
H(X|Y):=  - \sum_{j}
\sum_{i} p(x_i,y_j) \log  p(x_i|y_j)= -\sum_{j} q(y_j) \sum_{i}p(x_i|y_j) \log  p(x_i|y_j).
\end{equation}
Thus, we average the uncertainty about $X$ for each specific value of $Y$ over the distribution of those values. 
In short, the conditional entropy $H(X|Y)$ measures the amount of uncertainty in
$X$
given the knowledge of $Y$.\\
In our context, when we already know $Y$, that is, have obtained   the information $H(Y)$,  before receiving $X$, then $H(X|Y)$ is the reduction of uncertainty after having received $X$, or the information gained by learning $X$. Importantly, we want to assert here that \rf{5} is more fundamental than \rf{1}, because whatever we learn is conditioned on what we know already.  \\
Another important notion is the {\em mutual information} between $X$
and
$Y$ given by
\ba
\label{6}
MI(X:Y):&=& H(X) - H(X|Y) = H(Y) - H(Y|X) \\
\label{7}
&=& \sum_{i} \sum_{j} p(x_i,y_j) \log  \frac{p(x_i,y_j)}{p(x_i)
q(y_j)}.
\ea
The mutual information equals the reduction in the uncertainty
about
$X$ due to the knowledge of $Y$. If we do not know $Y$, our uncertainty about $X$ is $H(X)$, but as explained, when we know $Y$, the remaining uncertainty is only $H(X|Y)$, and the difference is mutual information. As \rf{6} indicates, and as follows from \rf{7}, it is symmetric in $X$ and $Y$. It is also non-negative, that is, by learning something, we cannot increase our uncertainty. \\
Before proceeding, we summarize the preceding:
\bel{8}
H(X)= MI(X:Y)+H(X|Y),
\qe
that is, 
$H(X)$, which quantifies how much you learn from observing $X$, is decomposed into 
 how much you learn about $X$ by observing $Y$, which is $MI(X:Y)$, and 
 how much you learn from observing $X$ when you already know $Y$, which is $H(X|Y)$. This is an instantion of Shannon's basic additivity principle described above.\\

For our subsequent analysis, we need to enhance the picture and also consider a 
 conditioning process with another random variable $Z$. Thus, we define  the {\em conditional mutual information}
\bel{9}
MI(X:Y|Z)=H(X|Z)-H(X|Y,Z).
\qe
This is the  mutual information between $X$ and $Y$ that can be gained with knowledge of $Z$. As already emphasized there always is some background knowledge on which we should condition. Usually, this is assumed implicitly, but for our discussion below, this needs to be made explicit. That is why we need \rf{9}. The formalism is so general that the actual role of $Z$ can vary. It could stand for the receiver's own memory, but it could also be some context information.\\
The standard {\em example} is the {\bf XOR} function where we have two random variables $Y,Z$ which assume their two values independently with probability $1/2$ each and which together determine the value of the random variable $X$ according to the following table\\

\hspace*{4cm} 
\begin{tabular}[h]{c c c}
$y$ & $z$ & $x$ \\
\hline
0  &0  &0    \\
1  &0  &1   \\
0  &1  &1 \\
1  &1  &0 
\end{tabular}\\

\noindent We have  $MI(X:Y)=MI(X:Z)=MI(Y:Z)=0$, but $MI(X:Y|Z)=MI(X:Z|Y)=MI(Y:Z|X)=1$, because knowing the values of two of the variables determines that of the third. 
This example shows that while  we always have  $H(X|Y)\le H(X)$, we do  not necessarily have $MI(X:Y|Z)\le MI(X:Y)$. That is, knowledge of the variable $Z$ may increase what we can learn from $Y$ about $X$, that is, make the knowledge of $Y$ more useful for deriving $X$. In this example, the knowledge of either $Y$ or $Z$ alone is completely useless for $X$, but together, they determine $X$ completely. This is a case of {\em complementary information}. In our interpretation below, $Z$ will stand for the context and what we can infer from $Y$ about $X$ will depend on that context. \\
This will be important for us, and we therefore need to describe the theory of {\em information decomposition} \cite{Liz}  that deals with this issue. A framework for information decompositions were first proposed in \cite{WB}, but since one may propose several, but partly conflicting axioms,  different versions have been developed, see for instance \cite{WB,HSP,GK,Gri1,JEC,QHS,Ince,CP,FL,Kol}. Here, I shall discuss the theory of \cite{BROJA} which is currently relatively widely accepted.  \\
As before, we have 
three random variables $X,Y,Z$. The  mutual information $MI(X:Y,Z)$ tells us how much  information  is gained about $X$ when we know $Y$ and $Z$ together. 
The question is how  $Y$ and $Z$ individually and jointly  contribute to this
information. 
The theory of information decomposition stipulates  four contributions to $MI(X:Y,Z)$:
\begin{eqnarray}
\nonumber
MI(X:Y,Z) =& & SI(X:Y,Z)\quad \text{shared information}\\
\nonumber
& +&U I(X: Y\backslash Z)\quad \text{unique information of Y}\\
\nonumber
& +&U I(X:Z\backslash Y )\quad \text{unique information of Z}\\
\nonumber& +& CI(X:Y,Z)\quad \text{complementary or synergistic
              information}.\\
  \label{10}
\end{eqnarray}
$U I(X: Y\backslash Z)$ is the information that $Y$ has exclusively, $SI(X:Y,Z)$ is the information that both of them have individually. $CI(X:Y,Z)$ is the information that can only be seen from the combination of $Y$ and $Z$, but from neither of them alone. Thus, the {\bf XOR} example has only synergistic information. \\
Another  example is  
{\bf AND} where again $Y,Z$  assume their two values independently with probability $1/2$ each and together determine the value of  $X$ according to the following table\\

\hspace*{4cm} \begin{tabular}{c c c}
$y$ & $z$ & $x$ \\
\hline
0  &0  &0   \\
1  &0  &0   \\
0  &1  &0   \\
1  &1  &1  
\end{tabular}\quad 

\vspace{.5cm}

\noindent $y$ and $z$ jointly determine $x$. 
When they both have the value  0, they share the information that $x=0$. Conversely, when $x=1$, then both $y$ and $z$ must also have the value 1, but when $x=0$, $y$ and $z$ cannot be fully recovered from $x$.  The mechanism loses some information. We have 
\be
\nonumber
H(Y,Z) =2 \text{ bits},
\qe
because there are four pairs of values of $Y$ and $Z$ with probability $1/4$ each, 
but
\be
\nonumber
H(X)=MI(X:Y,Z)=- \frac{1}{4}\log \frac{1}{4} - \frac{3}{4}\log \frac{3}{4}\approx .811 \text{ bits},
\qe
because $X$ assumes the value 1 with probability $1/4$ and 0 with $3/4$. Also, $H(X|Y)=1/2$, because while for $y=0$, we also have $x=0$, but if $y=1$, then $x$ can be either 0 or 1 with equal probability. Thus $MI(X:Y)=H(X)-H(X|Y)=.311$, and since the situation is symmetric in $Y$ and $Z$, also $MI(X:Z)=.311$.\\

As proposed in \cite{BROJA}, unique and shared information should only depend on the
marginal distribution of the pairs $(X,Y)$  and $(X,Z)$. This idea can be explained from an operational
interpretation of unique information: When $Y$ has unique information about $X$, then $Y$ should be able  to exploit this information, i.e., there should  be a setting where $Y$ can use this information to perform better than $Z$ at predicting $X$. $Y$ possesses unique information about $X$, if there exists a reward function for which $Y$ can achieve a higher expected reward based on the value $y$ and  knowledge of the conditional distribution $p(x|y)$ than by using instead the conditional distribution of $p(x|z)$. According to the scheme of \cite{BROJA}, to determine the unique information of $Y$ about $X$, we thus seek the distribution $Q$ with the same pairwise marginals  that minimizes $MI_Q(X:Y|X)$. For {\bf  AND}, this distribution  is  given by

\hspace*{4cm} \begin{tabular}{c c c|c}
$y$ & $z$ & $x$ & $p(y,z,x)$\\
\hline
0  &0  &0   &  1/2 \\
1  &1  &0   &   1/4 \\
1  &1  &1   &   1/4
\end{tabular}\quad 

\vspace{.5cm}

\noindent
For this $Q$, in fact $MI_Q(X:Y|Z)=0$ because now $Y$ and $Z$ are identical, instead of being independent. Thus, there is no unique information, and since $MI(X:Y)=U(X:Y|Z)+SI(X:Y,Z)$, we obtain $SI=.311$, and hence  $CI=.5$. Of course, if we just look at the pair $(Y,Z)$, then each possesses unique information as the other does not know its state. The transition from $(Y,Z)$ to $Y$AND$Z$ transforms unique into shared and complementary information.\\ 
In general, we may have both correlations between the input variables and relations created by the mechanism that computes $X$ \cite{RBOJB}. \\
As in this approach,  unique and shared information depend only on pairwise marginals, all higher order dependencies are contained in the synergistic information. Thereby, synergy becomes a measure of higher order interactions \cite{AOBJ,AJLS}. --  In this section, I have also used the presentation in \cite{ABJOR}.

\section{Transmission and computation}
The preceding provides formal concepts and foundations for two different issues that we should distinguish in a conceptual analysis. These are the transmission of information and the processing of information. According to Shannon's paradigm, information is transmitted from a sender to a receiver via a channel. Here, the usual aim is a faithful or lossless transmission. The receiver should receive and be able to decode everything that the sender has encoded and sent. The other aspect is the basic concept of computer science, the processing of information, or shorter, {\em computation}. We have discussed already the {\bf AND} and {\bf XOR} functions. Here, inputs are combined and utilized to generate an output. This is not lossless, because one cannot unambiguously recover the inputs from the output. In those two examples, the information $H(Y,Z)$ of the input variables was $\log  4=2$ bits, because we had four combinations, $(0,0), (1,0), (0,1), (1,1)$, occurring with equal probabilities, but the output information was only 1 bit for {\bf XOR} and $.811$ bits for {\bf AND}. Some information gets lost or forgotten on the way. But this is precisely the purpose of computation where one is interested in the output, and not in the inputs per se, but only insofar as they contribute to the output. \\
Perhaps in biology, these two aspects cannot be so neatly separated, but one might start on the premise that DNA inheritance involves transmission of information whereas DNA processing via transcription into RNA and translation into polypeptides is an instance of information processing, that is, some kind of computation. As we shall see, however, this needs to be modified.

\section{Carriers of information}
When information is transmitted across space, it is superimposed upon some carrier process, that is, the temporal domain is used. When information is transmitted across time, it is inscribed upon some material carrier and thus stored in space, that is, the spatial domain is used. This is admittedly somewhat rough, but it may guide a more refined analysis.\\
For the transmission of information, a medium or, in Shannon's terminology, a
channel is needed. This   could be simply physical space for visual signals,
it could  be the air for sounds, or it could be transmitted via
electromagnetic waves in  a telephone line, or whatever. Through such
channels, information is transmitted across space. When a message should be
transmitted across time, it is usually  inscribed into a more permanent
medium. This may  not be strictly necessary, and one might argue that oral, as
opposed to written, 
cultural traditions constitute  counterexamples. In most cases, however, a material carrier is used for the memorization and transmission of information into the future. Writing is an obvious technique for transmitting a message across time. Usually, symbols are written or engraved onto some neutral substrate, like paper, pergament, or stone. Such a substrate is not strictly necssary, as the Quipu script of the Inca  shows where knotted cords or  strings represent the symbols (although these strings are attached to some kind of twig as a backbone). \\
Digital information can be stored on very different carriers, and a central concept of computer science, the universal Turing machine, abstracts from the nature of such a substratum. In computer science, we have the distinction between hardware and software, and theoretically or ideally, a software can run on any hardware. (In practice, it seems that in modern technologies, this distinction gets often blurred, as programs are hardwired in computer chips or graphic cards.) The philosopher Hilary Putnam first developed \cite{Put1} and then rejected \cite{Put2} the thesis of functionalism that  meaning can be conceived independently of a substratum. The consciousness researcher Koch \cite{Koch}, for instance,  also rejects functionalism, as it would have the consequence, which he disputes,  that consciousness  does not need a physiological brain for its realization. While the question of functionalism has been mostly discussed in the cognitive sciences, the issue is obviously relevant for biology as well. But in biology, in contrast to computer science, we do not have a neat separation between hardware and software, and we should therefore be careful when transferring concepts from computer science to biology. But there are other and deeper reasons why functionalism is not adequate in biology, as will be discussed below.\\
Now in molecular biology, it is assumed that the DNA is the carrier of genetic information. The DNA is a string composed of instances of four different nucleotides, abbreviated as A,C,G and T. Such a nucleotide consists of a pentose, a phosphate, and a nitrogeneous base. The first two are the same for all nucleotides, and they link them in a sequence via phosphodiester bonds, whereas the nitrogeneous base (a purine for A and G or a pyrimidine for C and T) distinguishes the different nucleotides. The famous double helix is then formed by complementary bonds between those bases (A with T and C with G). Although most of the DNA is non-coding and has regulatory and structural roles (see for instance the conceptual discussion in \cite{SJ1,SJ2,JS1}), a triple of nucleotides can code for an amino acid (or represent a stop signal), the basic building block of a polypeptide, that is, of a constituent of a functional element, a protein. With very minor exceptions, that code is universal across all forms of life on earth. One might then consider the ribosome, the supramolecular complex of rRNAs and proteins, where protein synthesis takes place, as a system that transforms an input, a codon, that is, a triplet of RNA bases, into its output, an amino acid (or a synthesis stop, if the input is a stop codon). But of course, here the structure implements the genetic code, and we thus cannot physically separate hardware from software, although conceptually the computer science analogy may suggest that. We cannot simply reprogram the ribosome to utilize a different code for the translation of codons into amino acids, although in principle, this is not forbidden by the rules of chemistry.  We might also want to keep in mind here that both the input and the ribosome essentially consist of the same type of molecule, RNA. And the ribosomal proteins also get synthesized in the ribosome. Thus, we not only have the relatively trivial fact that in molecular biology, we do not find the conceptual computer science distinction between hardware and software realized, but more importantly, the distinction between program and input gets blurred. That latter distinction, in fact, is not confined to computer science, nor did it originate there. In mathematics, for a dynamical system, we have the distinction between the dynamical rule and its input. This gets a bit blurred, however,  when variable control parameters come into play. In physics, we have the  distinction between a physical law and its initial conditions. The question is to what extent an analogous distinction is also conceptually helpful in biology.\\
The DNA has a double role. On one hand, as described, it codes for the functional elements of the cell, the proteins.  The situation is not as simple as it might appear, as this takes place within  complex regulatory networks, as we shall discuss in more detail below. On the other hand, DNA copying is the basic mechanism for biological inheritance, although one could also say a lot about epigenetic inheritance \cite{JL,PSL}. \\
The question arises, however, to what extent we can legitimately speak of information here. As a first answer, let us try the following analogy, even though it will not be perfect. The past is the sender, the future is the receiver, and time is the channel. Thus, a solid structure like a piece of stone preserves its structure, and in that way, it also transmits information into the future. The future stone is not just a faithful copy of the past one, it is identical with it, except for some negligible wear. A man has to build a house for his family, but he does not give the house to his son, but only teaches him how to build a house himself. Thus, it is not the material structure that is transmitted (for instance, the house might only last for one generation and then collapse), but rather the knowledge for its construction. That knowledge might be rather concrete, like how to build a wall from wood or stones and mortar. It might also be more abstract, like how to recruit and supervise construction workers, or even more abstract, how to earn money to buy a house. An organism also transmits information beyond what is structurally embodied. It does not transmit  its full physical structure when it transmits information.  Thus, it needs to induce the creation of a new structure. Its own structural embodiment is, however, utilized for that transmission. That structural embodiment is the phenotype. Thus, on one hand, the genotype contains instructions on how to control and utilize external processes (a point to which we shall have to return) to construct its phenotype, and that phenotype then in turn enables the reproduction of the genotype. The essential information is contained in the genotype, but in order to transmit the DNA that carries the genotype, the phenotype of the organism is needed.  Still, there are two important problems.
\begin{enumerate}
\item Life is not a structure, but a process. Therefore, we should not think in terms of transmission of a structure, but rather in terms of the continuation of a process.
\item External processes are needed and utilized to enable the transmission of information. Biological information transmission is never a self-sufficient process. It depends on a complex environment. 
\end{enumerate}
The second point already has the consequence that functionalism, as discussed above, cannot be applied in biology. When we change the physical realization of a process, we also change its embedding into an external context. And since, as argued, this is fundamental for biological processes (or, for that matter, for cognitive processes as well), we cannot understand such a process properly in intrinsic terms, as functionalism suggests, but only by analyzing its complementarity with its context and its interaction with its environment. Importantly, that interaction may go in both directions. It is not the case that a biological organism is simply the passive recipient of external information. It rather shapes its environment actively, as emphasized by von Uexk\"ull \cite{Uex} and more recently conceptually combined with internal network regulation in \cite{LR}. In \cite{KPN}, the effect on the organism itself of its shaping of the environment, that is, how much information it transmits to itself through its environment,  is quantified in information theoretical terms (see also the discussion in \cite{J4}).

\section{Biological information}
Let us start with  considering the ribosome as an information processing machinery, a physical object that carries out a fixed program that handles variable input. Its input, a sequence of codons in an mRNA, contains the information which sequence of amino acids to produce. The program carrying out this processing is the universal genetic code. It seems that there is nothing conceptually wrong with this picture. Unfortunately, however, the biological reality is more complex than this, and therefore, this picture may be too simplistic and perhaps even misleading. 

Before explaining this, let us first consider another analogy, that of the
information transmitting role of the DNA as the essential feature of
biological inheritance. Here, the analogy seems to be that of a cooking
recipe, an assembly instruction or a blueprint for the construction of some
machinery. The DNA specifies, as in the recipe or the instruction, which
ingredients or parts to take to build up an organism. More precisely, it is an
instruction how to manufacture those ingredients or parts. The assumption
seems to be that the raw materials, the appropriate chemical molecules, as
well as  the necessary energy, are available and just need to be utilized and
processed. In some sense, this is true. A virus constitutes an example that
and  how external processes are utilized to enable the transmission of its
information. It modifies the operation of the host cell in order to transmit
the information for virus assembly. That is, the host cell becomes   the
machine that reproduces the virus information. From that perspective, any
biological organism does something analogous. It simply  induces its
environment to carry out its reproduction. One might recall  Uexk\"ull's concept \cite{Uex} of the ``Wirkwelt'' of an organism. 

Any organism thus modifies its environment to trigger its reproduction. From that perspective, the important information is not the assembly instruction from available material, but rather how to make that material available and let it operate, as the virus lets the host cell operate for it. As I am argueing in \cite{J10}, the key aspect of life is the control of processes, and here we have an instantiation of that principle. We can also consider this as an example of embodiment, a concept developed in theoretical robotics  \cite{PB,GZA,MGZA}. That means that the system in question, be it a biological organism, a cognitive system, or a robot, does not compute itself all the steps necessary for carrying out some task, but simply harnesses external regularities or processes for executing the task. The simplest example are physical laws like gravity and inertia which can complete a movement once initiated. No further computation from the side of the system is needed. Of course, much more complex processes can be enslaved for the system's purposes, like the host cell for the virus. Importantly, the system need not store the information about how that enslaved process operates. It only needs the, perhaps quite small, information about how to make the process work for itself. Thus, a viral genome is very small. The control of processes as the fundamental principle of life is discussed in more detail in \cite{J7,J10}.

But we need to think further. As I also argue in \cite{J10}, biological processes control each other. In particular, the information for initiating, carrying out, regulating or controlling  some process partly comes from that process itself, perhaps from its memory, but also partly from other processes. 

In molecular biology, there is the classical distinction between {\em cis} and {\em trans} information. When executing some genetic information, beyond the cis information contained in the genomic region in question itself, also trans information from other parts of the genome or even from outside the genome is needed to initiate and regulate the process. More generally, starting with the {\em operon} principle of Jacob and Monod \cite{JM}, it became clear that a complex regulatory machinery is superimposed upon the coding information contained in genomic regions, both at DNA and at RNA level (for a systematic analysis, see \cite{SJ1,SJ2,JS1}). From the perspective and with the tools described in Section \ref{info}, we can analyze the coordination emerging from the interplay between these two sources of information. Biologically, it is not simply the question of combining content information, coming from the genome, and processing information, coming from outside the genomic region in question, but the situation is much more intricate. For instance, the phenomenon of RNA splicing shows that the two only determine together what is produced, because one and the same genomic region is capable of producing different {\em genes} (in the terminology of \cite{SJ1,SJ2} that tries to clarify a bad conceptual confusion in modern molecular biology). And a sequence of nucleotides, at DNA or RNA level, not only contains coding information, but at the same time regulatory information, as it specifies which proteins or other RNAs can bind to it. In \cite{SJ1,SJ2}, we have called that regulatory information the {\em genon}. There is even more. The topological or geometric arrangement  of the DNA or other molecules in the cell plays an important role for the processing, see for instance \cite{Boi1}, and Klaus Scherrer and myself have coined the term {\em topon} to capture that spatial information.

We always need to carefully specify where and about what there is uncertainty and which factors contribute what, individually or jointly, to reducing that uncertainty. When we look at a genomic region, we can ask which and how many functional products will be produced from it. That depends not only on the regions coding information, but also on the superimposed regulatory information and on external regularity effects, coming from more or less specific protein and/or RNA combinations which in turn are affected by general external contexts  like temperature or gravity and specific external signals. When we look at a functional product, we may ask in turn what contributed to its production. The required analysis is different from the former.

Likewise when we look at genetic transmission, we can ask what kind of organism will emerge from a fertilized egg with the genetic information encoded in its DNA as well as the epigenetic information contained in that egg and mainly passed on from the mother. But the result depends not only, and in fact, only to a very small extent, on the genetic coding information. A much greater role is played by external factors that are needed to enable the development and growth of the organism by providing the necessary material, energy and context. Speaking metaphorically, the DNA takes this for granted, and formally speaking, we need to condition on all that information if we want to assess the role of the transmitted DNA. Unfortunately, this crucial point is often overlooked. \\
In our information theoretic formalism, this is easily formalized. We take the
random variable $Y$ for the transmitted genome, $Z$ for the external
contributions, and $X$ for the emerging new organism(s). According to \rf{10},
$Y$ then has unique ($UI$),
shared ($SI$), and complementary information ($CI$) about $X$ where what is complementary
depends on $Z$. But shared information  is superfluous from the perspective of
$Y$, because information that the environment can provide need not be encoded
in the genome. $Y$ may possess some unique information, but there seems to be
relatively little in biological organisms that develops autonomously and completely  independently of their environment.  We therefore postulate that the main evolutionary tendency is to increase the complementary information, in order to make best use of what the environment can potentially provide. If this is correct, then this also means that we cannot evaluate most of the genetic information without knowing the environment in which the new organism will develop, live and reproduce.

\section{Conclusion}
One can, of course, say that the DNA codes information about genes, that is, about the composition of functional products assembled by the molecular machinery of the cell. But, arguably, it is more important to know which functional products are produced when, under which circumstances, in which cells of a multicellular organism, in which combinations, and in what quantity. The information needed for that is complementary to the coding information. Some of it comes from other regions in the DNA that provide binding sites for regulatory proteins or simply organize the spatial structure of the genome. These regulatory proteins in turn are coded for by other regions in the DNA. Most of the regulation, however, takes place at RNA level, that is, after transcription \cite{Sch}. Here in particular, regulatory information is superimposed upon the coding information. In addition, signals from in- or outside the cell contain regulatory information. Again, that is complementary to the coding and regulatory information contained in the DNA or RNA strings. Those signals may determine which proteins are bound to or released from their binding sites \cite{JS1}. The methods of information decomposition described in Section \ref{info} may then complement the analysis of \cite{SJ2} for quantifying the various contributions to the information needed for producing a specific combination of functional products in a given cell. Importantly, this is meaningful only at a specific instance in the process. The different processes going on simultaneously or sequentially in a cell are coupled via many feedback loops, and react to  external signals, and in turn may produce such signals. Therefore, it is not possible to speak of some flow of causality along which information accumulates, even though it is possible in principle to analyze the various steps of some process with information theoretic tools \cite{SJ2}.  \\
The same applies at higher levels, in particular for organisms interacting
with their environments. There, we have discussed the concept of
embodiment. From that perspective, in \cite{GZLA}, the complementarity between internal and external
information is discussed in a framework called morphological computation. In \cite{KBOFA}, such principles have been invoked to analyze the concept of an individuum.\\
In either case, the key point is the complementarity of the information contained in the controlled process and that needed to control it. Only the latter is relevant for the controlling process. It does not need to know why and how the controlled process works \cite{J10}. \\
Importantly, applying information theoretical concepts and considerations in biology and elsewhere can only generate insight and avoid confusion, when it is clarified where and about what there is uncertainty that can be reduced upon receiving some signal. The nature of the signal, and in particular whether it depends on some material substrate or simply modulates some physical process, does not matter. What matters is the knowledge that the receiver possesses, for instance encoded in its memory, or  provided by other sources, for instance some external context or some regulatory scheme. When that is not carefully specified, applications of information theoretical concepts and considerations can be very misleading. \\
We have argued that functionalism is not an adequate theory for describing and
analyzing biological organisms. Biological life cannot be independent of its
material realization, because it is intricately intertwined with its
environment. But what is then the alternative? Should we resort to some kind
of Quniean holism where everything is directly or indirectly interwoven with
everything else? But then we would only be left with the biosphere of which
each organism is some part. This cannot be the solution either, as it would
not account for biological individuals. Biological organisms not only live in
their habitat or niche, but they control its processes to the extent needed
for their maintenance and reproduction, and perhaps even beyond that, and they
typically actively construct it. And internally, they develop regulatory
networks for the hierarchical or mutual control of processes. We have
developed information theoretical tools to analyze and quantify that
interdependence, both of an organism with its environment and of the internal
regulation. \\

{\bf Acknowledgment:} I thank two anonymous referees for constructive
comments.

\end{document}